\documentstyle[12pt,aaspp4]{article}

\begin{document}

\title {A circumstellar dust disk around T Tau N: \\
Sub-arcsecond imaging at $\lambda$ = 3 mm}

\author {R.L. Akeson,\altaffilmark{1}
\altaffiltext{1}
{Radio Astronomical Laboratory, UC Berkeley, Berkeley, CA  94720}
D.W. Koerner,\altaffilmark{2}
\altaffiltext{2} {University of Pennsylvania,
Dept.\ of Physics \& Astronomy, 209 S. 33rd St.,
Philadelphia, PA 19104-6396}
\& E.L.N. Jensen\altaffilmark{3} 
\altaffiltext{3}
{Dept.\ of Physics \& Astronomy, Arizona State University, 
Tempe, AZ 85287-1504}}

\begin{abstract}

We present high-resolution imaging of the young binary, T Tauri, in
continuum emission at $\lambda$=3~mm.  Compact dust emission with
integrated flux density 50 $\pm$ 6 mJy is resolved in an aperture
synthesis map at 0\farcs5 resolution and is centered at the position
of the optically visible component, T Tau N\null.  No emission above a
3$\sigma$ level of 9 mJy is detected 0\farcs7 south of T Tau N at the
position of the infrared companion, T~Tau~S\null.  We interpret the
continuum detection as arising from a circumstellar disk around
T~Tau~N and estimate its properties by fitting a flat-disk model to
visibilities at $\lambda$ = 1 and 3~mm and to the flux density at
$\lambda$ = 7~mm.  Given the data, probability distributions are
calculated for values of the free parameters, including the temperature,
density, dust opacity, and the disk outer radius.  The radial variation
in temperature and density is not narrowly constrained by the
data. The most likely value of the frequency dependence of the dust
opacity, $\beta$ = $0.53^{+0.27}_{-0.17}$, is consistent with that of
disks around other single T Tauri stars in which grain growth is
believed to have taken place. The outer radius, R =
41$^{+26}_{-14}$~AU, is smaller than the projected separation between
T Tau N and S, and may indicate tidal or resonance truncation of the
disk by T~Tau S\null.  The total mass estimated for the disk,
log(M$_D$/M$_\odot$) = ${-2.4}^{+0.7}_{-0.6}$, is similar to masses
observed around many single pre--main-sequence sources and, within the
uncertainties, is similar to the minimum nebular mass required to form
a planetary system like our own. This observation strongly suggests
that the presence of a binary companion does not rule out the
possibility of formation of a sizeable planetary system.

{\it Subject headings:} circumstellar matter --- stars:pre-main sequence ---
star:individual (T Tauri)
 
\end{abstract}

\section {Introduction}

Circumstellar material, generally believed to be in the form of disks,
is detected around most T~Tauri stars (Strom et al.\ 1989; Beckwith et
al.\ 1990). A comparison of inferred disk properties with theoretical
constructions suggests that these disks play a dual role as a source
of accretion material for star formation (Shu et al.\ 1993) and a
reservoir of material from which planetary systems may arise (cf.\
Beckwith \& Sargent 1996; Koerner 1997).  This picture is complicated,
however, by results of recent surveys: most T~Tauri stars are in
binary or multiple systems (Ghez et al.\ 1993; Leinert et al.\ 1993).
On average, millimeter-wave flux densities are reduced for T~Tauri
binaries with separations between 1 and 100~AU as compared to T~Tauri
stars in single or wide binary systems (Jensen et al.\ 1994; 1996;
Osterloh \& Beckwith 1995). This range of separations largely
encompasses the distribution of radii observed for dust disks around
young, low-mass stars and emphasizes the likelihood of a direct impact
of companions on circumstellar disks.  By determining the spatial
distribution of dust in binary systems with separations of several
tens to 100~AU, the nature of disk/companion interactions can be
investigated empirically.

T Tauri was identified initially as a prototype of a class of
pre-main-sequence stars (cf.\ Bertout 1989 and references therein) and
discovered to be a binary in the infrared
(Dyck et al.\ 1982).  The
optically-obscured companion, T~Tau~S, is located 0\farcs7 south of
the visible star T~Tau~N, corresponding to a projected separation of
100~AU at the 140~pc distance of the Taurus-Auriga dark cloud
(Wichmann et al.\ 1997). Both sources are detected at 2 and 6 cm
accompanied by a bridge of connecting emission (Schwartz et al.\
1986). Together with proper motion studies (Ghez et al.\ 1995), the
morphology of this emission strongly suggests that the system is
physically bound. T Tau is also associated with a substantial mass of
dust, as evidenced by polarized nebulosity at infrared wavelengths
(Weintraub et al.\ 1992) and by one of the highest $\lambda$ =
1 mm luminosities among young stars (Adams et al.\ 1990;
Beckwith et al.\ 1990).

Although T~Tau~N is at least 7 magnitudes brighter than T~Tau~S at
optical wavelengths (Stapelfeldt et al.\ 1998), the infrared
luminosity of the system is dominated by emission from T~Tau~S (Ghez
et al.\ 1991).  Similar instances of differential extinction are found
in a handful of other ``Infrared Companion'' (IRC) systems and have
led to speculation about the precise nature of their origin.
Competing IRC models invoke different roles for the interaction
between the companions and the distribution of surrounding dust and
gas. Observations of the thermal dust emission at millimeter
wavelengths can be used to determine the mass distribution.  Nearly
all such observations carried out thus far, however, do not have the
spatial resolution needed to discriminate between a circumbinary disk
or envelope and circumstellar disks around either component.  Both
components of T Tau are thought to have circumstellar disks based on
their infrared excesses, but these observations are sensitive primarily to
temperature and do not provide a good measure of the dust {\it mass}
distribution.  The highest resolution mm-wave image available to date
was obtained recently by Hogerheidje et al.\ (1997) and suggests that
the 1~mm continuum emission is dominated by radiation from T~Tau~N,
although the synthesized beam in their map (FWHM 0\farcs8) is slightly
larger than the binary separation.

To ascertain the respective masses of any circumstellar disks around
the components of T Tau, long-wavelength measurements of optically
thin dust emission must be carried out at a spatial resolution that
unambiguously separates the respective contributions of the binary
components. Observations at $\lambda$ = 3~mm probe the mass of the
densest material and are ideal for distinguishing compact
circumstellar dust from that of a surrounding envelope.  They are also
important in distinguishing between free-free and thermal dust
emission.  Accordingly, we have used the new long baseline capability
of the BIMA mm-array\footnote{The BIMA array is operated by the
Berkeley Illinois Maryland Association with funding from the National
Science Foundation.} to make high resolution 3~mm continuum
observations of T~Tau.

\section{Observations and Results}

Observations of T~Tau at $\lambda$ = 2.8 mm were carried out from 1996
December to 1997 March with the 9-element BIMA array.  Continuum
emission was measured in an 800 MHz bandwidth centered at 108 GHz.
Data were taken with the array in the A, B, and C configurations,
providing baseline coverage from 2.1 k$\lambda$ to 420 k$\lambda$ with
sensitivity to emission on size scales up to $\sim$60\arcsec.
Interleaved observations of a nearby quasar, 0431+206, were included
as a check on the phase de-correlation on long baselines.  The phase
calibrator was 0530+135, with an assumed flux of 3.1~Jy during A
array.  Data were calibrated and mapped using the Miriad package.  A
map of 0431+206 yielded an image of a point source, indicating little
atmospheric degradation of the resolution in the observations of
T~Tau.

\begin{figure}
\plotfiddle{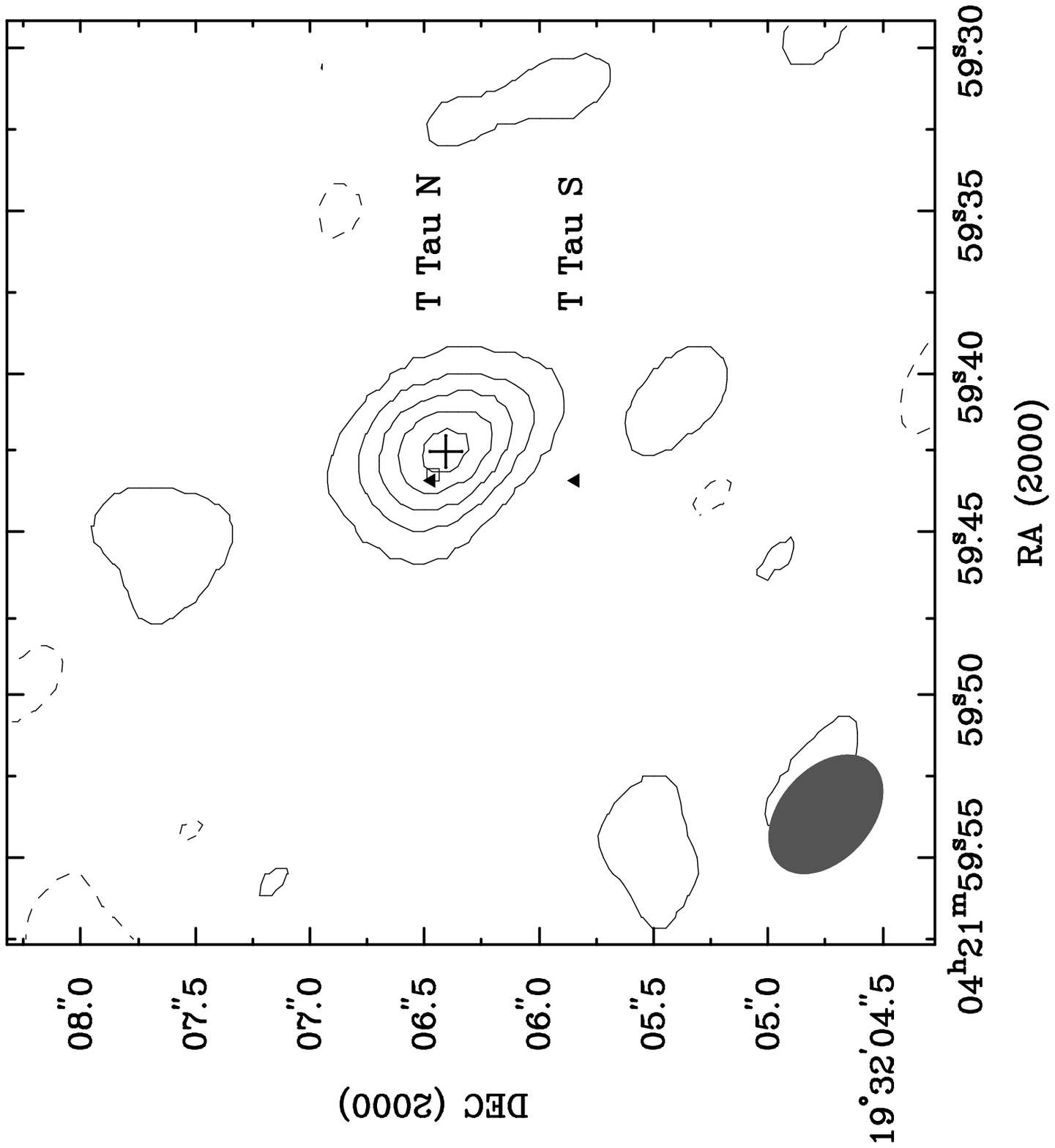}{6in}{270}{90}{90}{-350}{500}
\caption{The 108~GHz continuum emission using only the longest spacing
array.  The beam size is 0\farcs59 $\times$ 0\farcs39 at a position
angle of 48\arcdeg.  The peak flux is 32 mJy/beam and the RMS is 3.1
mJy/beam.  The contour levels plotted are 2$\sigma$.  The size of the
error bars on the millimeter emission center position (cross)
represents the absolute positional uncertainty for the millimeter
image of 0\farcs07.  The filled triangles mark the radio positions of
T~Tau~N and S from Skinner \& Brown (1994), where the error bars
(which do not include any possible systematic errors) are within the
symbol.  The open square marks the optical position of T~Tau~N from
Hipparcos.  All positions were adjusted to epoch 1997.0 using a proper
motion of $\mu_{\alpha cos \delta}=0\farcs015$ yr$^{-1}$ and
$\mu_{\delta}=-0\farcs012$ yr$^{-1}$ (Hipparcos).  }
\label{map}
\end{figure}

An aperture synthesis image of T Tau was constructed with data from
the A array alone (the longest baselines) and is displayed in Figure
\ref{map}.  The beam size is $0\farcs59 \times 0\farcs39$ at a 
position angle of 48\arcdeg.
It is clearly evident that the compact 3~mm emission
arises from circumstellar material surrounding T~Tau~N only.  Peak
emission of 32 mJy beam$^{-1}$ is centered at RA(J2000) 04:21:59.424
and Dec(J2000) 19:32:06.41 with an absolute positional uncertainty of
$\sigma$ = $\pm$0\farcs07 as determined from observations of
0431+206.  The peak position is within 1.6$\sigma$ of the Hipparcos
coordinates for T~Tau~N, but 7.8$\sigma$ distant from the position of
T~Tau~S\null.  The emission is resolved with an approximate
deconvolved source size of 0\farcs45 $\times$ 0\farcs32 (FWHM) (63
$\times$ 45~AU) at PA 19\arcdeg, assuming an elliptical Gaussian
shape.  The integrated flux density, 50 $\pm$ 6 mJy (where the
uncertainty is dominated by the flux calibration error), is consistent
with the 100 GHz value measured at lower resolution by Momose et al.\
(1996), 48 $\pm$ 7 mJy.

No emission is detected at the position of T~Tau~S above the 3$\sigma$
upper limit of 9 mJy beam$^{-1}$, and no circumbinary emission is
apparent.  The integrated flux density detected in the compact C array
is no greater than that for A array. Since the C array observations
are most sensitive to emission on larger spatial scales, up to 60$''$,
the absence of detectable excess emission implies that the majority of
continuum emission detected in a 1$'$ aperture at $\lambda$ = 3 mm
originates from a circumstellar region around T Tau N\null.  The
3$\sigma$ upper limit on circumbinary emission, accounting for
thermal noise and flux calibration errors, is 17 mJy.

\begin{figure}
\plotfiddle{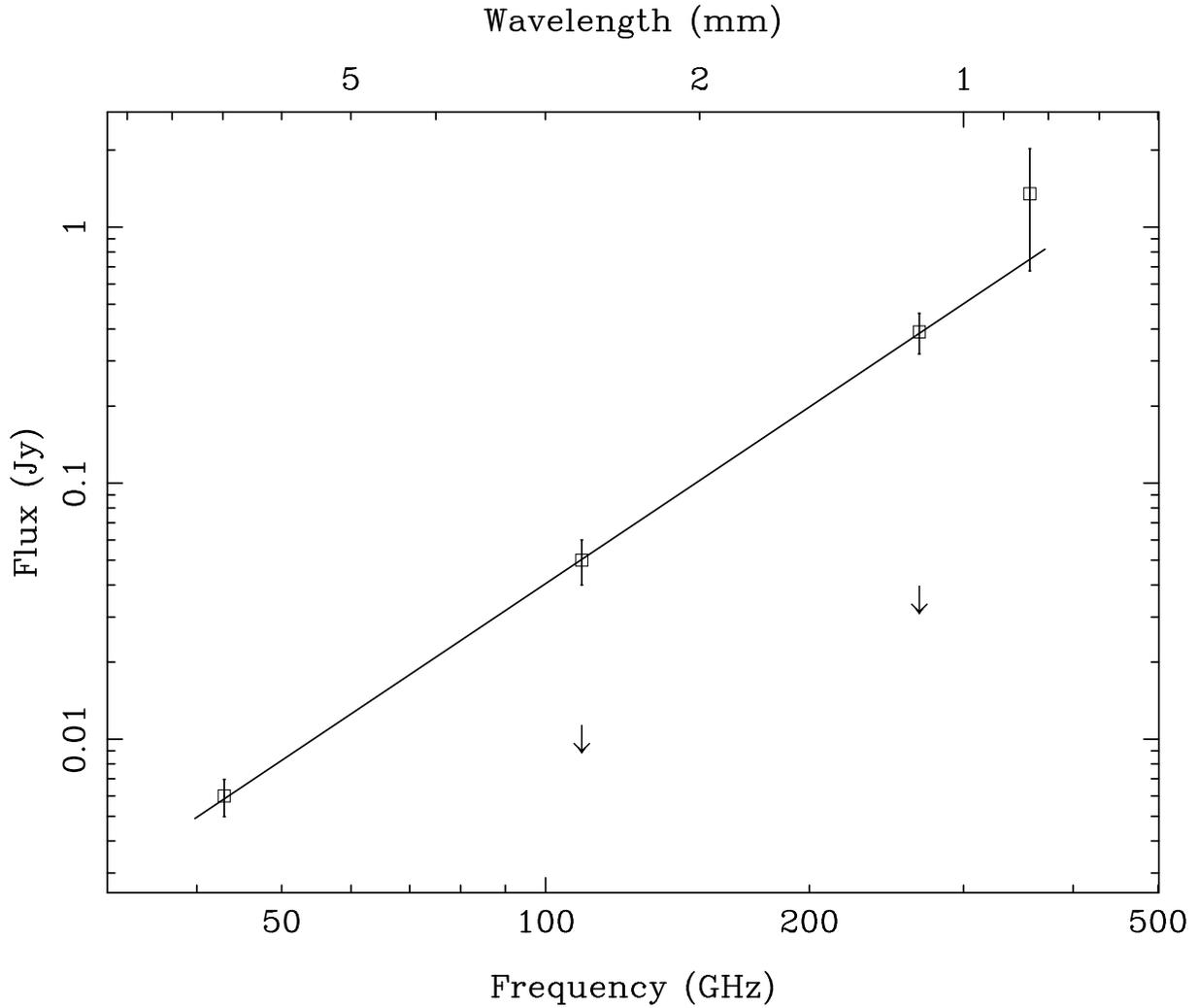}
{5.0in}{270}{90}{90}{-345}{475}
\caption{Millimeter and sub-millimeter fluxes for T~Tau~N (squares)
and 3 $\sigma$ upper limits (arrows) for T~Tau~S\null.  The 267 GHz
($\lambda$=1 mm) and 357 GHz ($\lambda$=0.8 mm) fluxes are from
Hogerheidje et al.\ (1997).  The 43 GHz ($\lambda$=7 mm) flux is from
Koerner et al.\ (1998).  The solid line fit is plotted with
$\alpha$=d(log($F_\nu$))/d(log($\nu$)) of 2.3.
\label{sed}}
\end{figure}

Our measurement of the 108 GHz flux density from T~Tau~N is plotted in
Figure \ref{sed} together with high-resolution measurements at 43 GHz
($\lambda$ = 7~mm) (Koerner et al.\ 1998), 267 GHz ($\lambda$ = 1~mm)
and 357 GHz ($\lambda$ = 0.8~mm) (Hogerheidje et al.\ 1997).  All four
points can be fitted by a single power-law curve with $\chi^2$ = 1.8 and
goodness of fit 0.4.  The best-fit spectral index
$\alpha$=d(log($F_\nu$))/d(log($\nu$)) is 2.30$\pm$0.1.  This value is
in good agreement with those from disks around single T Tauri stars
and consistent with thermal emission from large circumstellar dust
grains (Beckwith \& Sargent 1991; Mannings \& Emerson 1994; Koerner et
al.\ 1995), suggesting that the continuum emission from T Tau N has a
single origin in thermal dust emission along the entire wavelength
range from $\lambda$ = 1 to 7~mm.

We argue that the 3mm emission arises from a circumstellar disk around
T Tau N\null.  The spectral slope is consistent with radiation from
dust grains (see Koerner et al.\ 1998 for a more stringent limit on
the possible contribution of free-free or gyrosynchrotron emission).
The star is detected optically, even though the lower limit to the
dust mass is high; if the dust were distributed in a uniform density
sphere with a size given by the A array fit, the extinction to the
star would be A$_V \sim$ 1000.

\section {Circumstellar disk models for T~Tau~N}

\subsection{Model description}

To refine estimates of the parameters of dust around T~Tau~N, we fit
the visibility amplitudes directly with a model of a circumstellar
disk.  While computationally intensive, fitting in the visibility
plane rather than the image plane avoids the non-linear process of
deconvolution and allows the instrumental errors to be included
in a consistent manner.  

Our disk model is thin and circularly symmetric with a flux from an annular
region at radius $r$ given by
\begin{equation}
dS_{\nu}(r) = {2 \pi \cos\theta \over D^2 } B_{\nu} (1 - e^{-\tau}) r \, dr,
\label{disk_flux}
\end{equation}
where $B_{\nu}$ is the Planck function, $D$ is the distance, and $\theta$
the inclination angle.  Flared disks have been invoked to explain
the flat infrared spectral indices of some T Tauri stars 
(Kenyon \& Hartmann 1987);
however, the effects due to flaring are not important at these 
long wavelengths (Chiang \& Goldreich 1997).
The visibility amplitude $V$ at $uv$-distance $\eta$ is calculated
with a Hankel transform,
\begin{equation}
V(\eta) = 2\pi \int S_{\nu}(r)J_0(2\pi\eta r/D)r\,dr,
\label{hankel}
\end{equation}
where $J_0$ is the Bessel function. The inner radius is set by the dust
destruction temperature (2000~K); the exact value has little effect on
the millimeter flux density.
The temperature, surface density and dust opacity are described by
power-law relations: $ T(r) = T_{\rm 10~AU} (r/{\rm 10~AU})^{-q},\
\Sigma (r) = \Sigma_{\rm 10~AU} (r/{\rm 10~AU})^{-p},$ and $\kappa_{\nu} = 
\kappa_o (\nu/\nu_o)^{\beta}$.  Due to its dependence on the
product of surface density and dust opacity, the optical depth scales
with the corresponding power-law exponents in radius and frequency,
\begin{equation}
\tau (r,\nu) = {\Sigma \kappa_{\nu} \over \cos\theta} \equiv
\tau_{\rm 10~AU} \Big({r \over {\rm 10~AU}}\Big)^{-p}\Big({\nu
\over \nu_o}\Big)^{\beta}.
\end{equation}
The reference value for $\kappa_o$ is 0.1~g$^{-1}$~cm$^{2}$ at
$\nu_o = 1200$ GHz ($\lambda$ = 250~$\mu$m; Hildebrand 1983).  

Eight parameters were varied in the model-data comparison: $T_{\rm
10~AU}$, $ q$, $\tau_{\rm 10~AU},$ $p,$ $\beta,$ $r_{out},$ $\theta$,
$\mbox{and}\ \alpha$, the position angle of the disk on the sky.  In
addition to the 3~mm visibilities, those measured at 1~mm by
Hogerheidje et al.\ (1997) were used together with the 7~mm flux
density (Koerner et al.\ 1998).  The IRAS 100 $\mu$m flux of 120 Jy
(Strom et al.\ 1989) was included as an upper limit.  For each value
of $\theta$ and $\alpha$, the $u$ and $v$ coordinates for each
visibility were de-projected to those of a face-on
disk, then binned in annuli of de-projected
$uv$-distance for comparison to model values calculated by
Eqn.\ \ref{hankel}. Over 5 million models were compared to the data 
within an 8-dimensional grid in parameter space. 
Logarithmic grid spacings were used for the
temperature, optical depth and outer radius.

To quantitatively assess the reliability of estimates of properties of
a disk around T Tau N, we calculate the probability distribution for
each parameter over the entire range of models.  A detailed description
of this Bayesian approach is given in Lay et al.\ (1997).  Given the
data, the probability of a model with a particular set of parameter
values is proportional to $e^{-\chi^2}$ where $\chi^2$ is the standard
squared difference between data and model, weighted by the uncertainty
in the data.  A systematic error in overall flux calibration was
accounted for by normal weighting of a range of model flux scalings
with 1$\sigma$ corresponding to a 10\% difference in flux.  
The final probability for a given model was taken
as the sum of probabilities over all flux scalings and multiplied by a
factor of $\sin \theta$ to account for the fact that edge-on disks are
more likely than face-on in a randomly oriented sample.  Finally, the
relative likelihood of each parameter value was calculated by adding
the probabilities of all models with that parameter value.

\subsection{Parameter results}

The resulting probability distributions for the disk parameters are
given below.  The ranges considered for $T_{\rm 10~AU}$, $\tau_{\rm
10~AU}$, and $r_{out}$ were sufficiently wide to bracket all values
with significant probability.  Parameters values quoted are at
the median of the probability distribution with an error range
encompassing 68\% of the total probability.  However, it is important
to keep in mind that the distribution is not Gaussian.

\begin{figure}
\plotfiddle{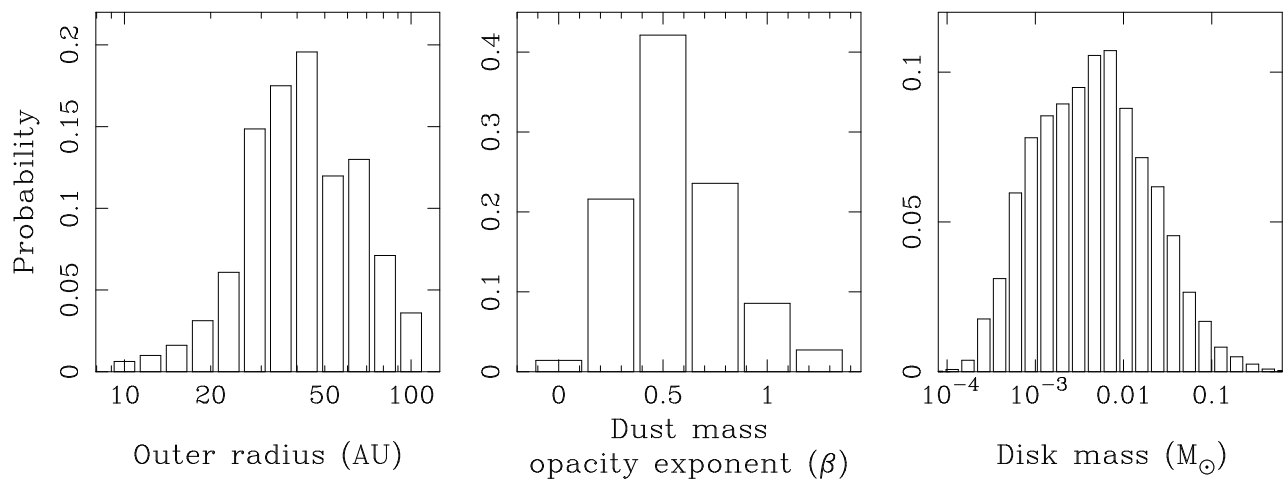}
{2.5in}{270}{70}{70}{-270}{280}
\caption{The probability distributions for the disk outer radius,
dust mass opacity exponent and derived disk mass.
The likelihood of each parameter value was calculated by summing
the probabilities for models with all possible values of the
remaining parameters.  The relative value of the probability
is more important than the absolute value, which depends on
the number of parameter values used in the calculations.
\label{prob}}
\end{figure}

The disk outer radius is $r_{out} = 41^{+26}_{-14}$~AU (Figure
\ref{prob}).  The probability that it exceeds the projected binary
separation (100 AU) is only 3\%.  Models with large outer radii
generally have higher values of $p$ and lower values of $\tau_{\rm
10~AU}$.  The value for the dust mass opacity index, $\beta =
0.53^{+0.27}_{-0.17}$ (Figure \ref{prob}), 
is consistent with values measured for
circumstellar disks around T Tauri stars (Beckwith \& Sargent 1991;
Koerner et al.\ 1995).  There is an 85\% probability that $\beta$ $\ge
0.30$, the value calculated by assuming optically thin emission
(S$_\nu \propto \nu^{2 + \beta}$) and fitting a straight line to the
7, 3, and 1~mm fluxes.  As discussed below, this implies that at least
some of the emission is optically thick.  There is some correlation
between models with high values of $\beta$ and those with high $\tau$
and low $T$.  The temperature, $T_{\rm 10~AU} = 26^{+34}_{-13}$~K, and
$\lambda$ = 3~mm optical depth, $\tau_{\rm 10~AU} =
0.50^{+0.83}_{-0.36}$, are not as tightly constrained as $r_{out}$.
Although many models have $\tau > 1$ at 10~AU, most (89\%) radiate
more than half their total emission in an optically thin regime.  Note
that for an optically thin disk in the Rayleigh-Jeans regime, Eqn.\
\ref{disk_flux} becomes degenerate in $T$ and $\tau$.  Consequently,
temperature and optical depth are anti-correlated for $\tau <$ 0.5 at
10~AU.  The data do not narrowly constrain values for $q$, $p$,
$\theta$ or $\alpha$.  The ranges used for $p$ and $q$ were
$p$=0.5--2.0 and $q$=0.4--0.75.  Steeper density profiles are slightly
favored over shallow ones; the probability that $p$ is $\ge$ 1.5 is
65\%.  The lack of preferred values for $p$ and $q$ is largely due to
the degeneracy of $p$ and $q$ for optically thin emission.

It may be possible to constrain the disk parameters further by including
data from additional wavelengths.  Mid-infrared flux densities are
often used to determine $q$ and $T_o$, for example.  We chose not to
include these data, however, because the mid-infrared flux traces
material within a few AU of the star, while the millimeter data is
sensitive mainly to material tens of AU away.  The simple power-law
relations assumed in the model may not be valid over such a large
range of disk radii and physical conditions.  

The disk model masses, weighted by the model probabilities, were
binned to derive a median value log(M$_D$/M$_\odot$) =
${-2.4}^{+0.7}_{-0.6}$ (Figure \ref{prob}). 
The wide range in mass is due largely to the
range of $\tau$ values that fit the data.  For a given 3~mm flux,
disks with higher mass correspond to those with higher $\tau$ and
$\beta$.  The median value for the disk mass, M$_D$ = $4 \times
10^{-3}$ M$_\odot$, is toward the low end of disk masses typically
derived for classical T~Tauri stars (e.g.\ Beckwith et al.\ 1990).  We note,
however, a large dependence of mass estimates on values assumed for
the other parameters.  Beckwith et al.\ assumed $\beta$ = 1 and an
outer radius of 100~AU\null.  If we consider only models with $\beta
\ge 0.75$ and an outer radius $>$ 50~AU, the median mass increases to
$10^{-2}$ M$_\odot$, similar to the minimum mass solar nebula.
 
\section{Discussion}

It has been conjectured that the formation of planetary systems arises
from the collapse of protostellar clouds which rotate more slowly than
clouds from which binaries form (Safronov \& Ruzmaikina 1985).  This,
in turn, raises the possibility that planetary systems may fail to
form in the binary environment.  In contrast, our observations and
modeling demonstrate that a substantial mass of material, with size
like that of the solar system, can exist around a star in a binary
system with separation not less than 100 AU\null. This material
constitutes a large reservoir available to planet-forming processes;
the low value of the mass opacity index, $\beta$ =0.53, further
suggests that the formation of larger grains may already be underway
as a first step toward planetesimal formation (Beckwith \& Sargent
1991; Mannings \& Emerson 1994; Koerner et al.\ 1995).

Our observations rule out a greater circumstellar mass of dust around
T Tau S---whether in an edge-on circumstellar disk or compact
spherical envelope---as a simple explanation for the origin of
increased extinction along the line of sight to T Tau's infrared
companion. Due largely to the low optical depth of dust continuum
emission at $\lambda$ = 3mm, however, the true source of extra
extinction is not identified unambiguously.  Our estimate of the outer
radius is smaller than the projected separation between T~Tau~N and
S and suggests that a disk around T Tau~N does not obscure
T~Tau~S\null.  However, we did not consider models with an exponential
density profile at the outer edge like that of Hogerheidje et al.\
(1997).  We note that since $\kappa(500\ {\rm nm})/\kappa(3\ \rm{mm})
\sim 10^3$--$10^4$ (e.g., Pollack et al.\ 1994), the material that
provides the extinction toward T Tau S could easily be undetectable at
3 mm.  Thus, our data are consistent either with obscuration of T Tau
S by tenuous outer regions of the T Tau N disk or with truncation of
the T Tau N disk by T Tau S as discussed below.

If the binary components are in a bound orbit, as suggested by their
common proper motion, the distribution of circumstellar material at
the outer edge of the disk will be affected by the gravitational
influence of the companion.  In models of disk/companion interactions,
the size of the circumstellar disk depends on the mass ratio of the
stars and their separation (Papaloizou \& Pringle 1977, Artymowicz \&
Lubow 1994).  The radius of the circumprimary disk typically ranges
from 0.3 to 0.4 times the separation for mass ratios of 1 to 0.3 and
circular orbits.  If the orbital plane of the system is viewed 
nearly face-on and 
the binary separation is 100--110~AU, the tidal
disk radius would be 30--40~AU\null.  The disk size estimated
from modeling our observations, 27 $< R <$ 67~AU, is consistent with
the range of values predicted for tidal truncation.  If tidal
truncation has occurred and we are seeing the true size of the disk in
our maps, then the disk around T Tau N is not obscuring T Tau
S\null. However, high-resolution observations at sub-millimeter
wavelengths or in molecular transitions that better trace
low-column-density material are needed to adequately solve this
problem.

Finally, we point out that the disk around T Tau N is similar to those
observed around some single low-mass stars, regardless of the reliability
of model assumptions that lead to an estimate of the absolute value of
its mass and size, since the millimeter-wave flux from T Tauri is
among the brightest measured from a large sample of young low-mass
stars (cf.\ Beckwith et al.\ 1990, Osterloh \& Beckwith 1995), and the
nominal FWHM size of the emission is similar to that of single-star
disks for which sufficiently high-resolution observations have been
carried out (e.g. Lay et al.\ 1994; Mundy et al.\ 1996). If the
largest disks around T Tauri stars typically yield planetary systems
like our own, it is plausible that the disk around T Tau N will too,
in spite of the presence of a companion at a distance of 100 AU or greater.  
Since the majority
of pre-main-sequence stars are in multiple systems, the possibility of
planetary formation in binaries like T Tauri invites consideration of
the likelihood that a non-negligible fraction of binary stars
contribute substantially to the estimated fraction of stars that
possess planetary systems.

\acknowledgments

We are grateful to M. Hogerheidje for allowing us to use the 1~mm
data.  We thank L. Looney and L. Mundy for help with observing and
O. Lay for helpful discussions.  RLA acknowledges support from the
Miller Institute for Basic Research in Science.  ELNJ and DWK gratefully
acknowledge support from the National Science Foundation's Life in
Extreme Environments Program through grant AST-9714246.

\end{document}